\documentclass[superscriptaddress, aps, prl, reprint, twocolumn, amsmath, amssymb, showpacs]{revtex4-1}
\usepackage{graphicx}  
\usepackage{mathptmx}
\usepackage{epstopdf}
\usepackage{dcolumn}   
\usepackage{bm}        
\usepackage{amssymb}   
\usepackage{amsmath}   
\usepackage{amsthm}
\usepackage{chemarrow}
\usepackage{color}
\usepackage{xcolor} 
\usepackage{mathrsfs}
\usepackage{float}
\usepackage{physics}
\usepackage{bbold}
\usepackage{bbm}
\usepackage[toc,page]{appendix}
\usepackage{longtable}
\usepackage{multirow}
\usepackage{upgreek}
\usepackage{epstopdf}
\usepackage{ulem}
\usepackage{graphicx}
\usepackage{lipsum}
\usepackage{subfigure}
\usepackage[]{ezedits}
\defineEdit{WJ}{\color[rgb]{0.5,0.3,0.9}}{\color[rgb]{0.5,0.3,0.9}}
\defineEdit{XCH}{\color[rgb]{0.9,0.45,0.0}}{\color[rgb]{0.9,0.45,0.0}}
\defineEdit{DB}{\color[rgb]{0,0.3,1}}{\color[rgb]{0,0.3,1}}
\defineEdit{LY}{\color[rgb]{0,0.3,1}}{\color[rgb]{0.7,0.7,0.3}}

\usepackage[a4paper, colorlinks=true, linkcolor=blue, citecolor=blue,
pdfauthor={ },
pdftitle={ },
pdfsubject={ },
pdfkeywords={ }]{hyperref}
\usepackage[version=4]{mhchem}
\usepackage{ulem,fancyvrb}

\usepackage{mathrsfs}

\theoremstyle{plain}
\newtheorem*{theorem*}{Theorem}

\begin{document}
	

\title{A tunable  feedback-controlled magnetic trap for a magnet in free fall}
\date{\today}
	\author{Changhao Xu}
	\affiliation{Johannes Gutenberg University, Mainz 55128, Germany}
	\affiliation{Helmholtz-Institute, GSI Helmholtzzentrum fur Schwerionenforschung, Mainz 55128, Germany}
    \author{Alexander Heidt}
	\affiliation{Leibniz University Hannover (LUH), Institute of Transport and Automation Technology, Hannover 30167, Germany} 
    \author{Mohammadreza Nematollahi }
     \affiliation{Johannes Gutenberg University, Mainz 55128, Germany}
	\affiliation{Helmholtz-Institute, GSI Helmholtzzentrum fur Schwerionenforschung, Mainz 55128, Germany}
    \author{Christoph Lotz}
	\affiliation{Leibniz University Hannover (LUH), Institute of Transport and Automation Technology, Hannover 30167, Germany} 
     \author{Ernst Maria Rasel}
	\affiliation{Leibniz University Hannover (LUH), Institute of Quantum Optics, Hannover 30167, Germany} 
    \author{Yan Liu}
    \email[Corresponding author: ]{liuyan01@uni-mainz.de}
	\affiliation{Helmholtz-Institute, GSI Helmholtzzentrum fur Schwerionenforschung, Mainz 55128, Germany}
    \affiliation{Technology and Engineering Center for Space Utilization, Chinese Academy of Sciences, Beijing 100094, China}
    \author{Wei Ji}
	\email[Corresponding author: ]{wei.ji@pku.edu.cn}
	\affiliation{School of Physics and State Key Laboratory of Nuclear Physics and Technology, Peking University, Beijing 100871, China}

\author{Dmitry Budker}
\email[]{budker@uni-mainz.de}
\affiliation{Johannes Gutenberg University, Mainz 55128, Germany}
\affiliation{Helmholtz-Institute, GSI Helmholtzzentrum fur Schwerionenforschung, Mainz 55128, Germany}
\affiliation{Department of Physics, University of California, Berkeley, California 94720, USA}

\begin{abstract} 
 
 Ferromagnets in free space are predicted to exhibit pure Larmor precession at near-zero magnetic fields and provide exceptional sensitivity for magnetometry and gyroscopy. Notably, pure Larmor precession has not been observed in a macroscopic ferromagnetic particle, despite its fundamental importance and potential for probing relativistic effects and dark-matter interactions. Realizing such dynamics requires true free fall to eliminate clamping losses and trap-induced systematics. A central challenge is designing a tunable trap that is weak enough to permit near-free evolution yet robust enough to withstand the disturbances of launch and release. Here, we propose and demonstrate a novel master proportional-integral-differential magnetic trap (MPIDMT) combining a PID-controlled coil system with a master control coil system.
 Implemented in the third-generation drop tower - Einstein-Elevator, during the microgravity phase the system stably levitates a ferromagnetic particle against shock accelerations up to 1.5\,g and resolves its motion in both a low-field (0.4\,g) configuration and in pure free fall. These results represent a key step toward free-fall ferromagnetic magnetometry, the long-sought direct observation of macroscopic Larmor precession, and future space-based experiments.

\end{abstract}

\maketitle

\textit{Introduction --}
Precision measurement of magnetic fields is crucial across a wide range of sciences, from life sciences—such as detecting brain activities and cardiac signals\,\cite{boto2018moving,cohen1972magnetoencephalography,aslam2023quantum}—to fundamental physics, including searches for phenomena beyond the Standard Model\,\cite{wei2023ultrasensitive,xu2024constraining,ji2018new,cong2025spin}. Ferromagnetic sensors have been predicted to achieve unprecedented sensitivity because the strong spin correlations within a ferromagnet can substantially suppress spin-projection noise, which is the fundamental noise source limiting atomic magnetometers\,\cite{kimball2016precessing,band2018dynamics,vinante2021surpassing,ni2025microscopic}. Moreover, a ferromagnetic magnetometer can operate under ambient conditions\,\cite{ji2025levitated}, unlike the highly-sensitive spin-exchange-relaxation-free (SERF) magnetometers that require stringent magnetic shielding, or superconducting quantum interferometer devices (SQUID) that must be cooled to cryogenic temperatures\,\cite{allred2002high,clarke2006squid}.

In order to achieve the optimum performance of the magnet, the ferromagnet must be mechanically isolated from its surroundings so that it can move freely without clamping or contact forces. On Earth, this necessitates levitation, achieved by counteracting gravity through techniques such as superconducting levitation\,\cite{wang2019dynamics,vinante2020ultralow,ahrens2024levitated}, diamagnetically stabilized magnetic levitation\,\cite{simon2001diamagnetically,ji2025levitated,leng2024measurement}, or magnetic Paul-trap levitation\,\cite{perdriat2023planar}. Each approach, however, introduces unwanted effects—such as residual magnetic fields or dynamical noise from the trap itself.

To reach the sensitivity limit of the magnet, it must ideally operate in a free-fall or near-free-fall state, where the trapping field and trap-induced noise are minimized\,\cite{kimball2016precessing}. The free-fall condition further enables the observation of the precession of a macroscopic particle whose angular momentum is dominated by quantum-mechanical spin\,\cite{kimball2016precessing,fadeev2021ferromagnetic,fadeev2021gravity}. Thus far, only indirect evidence of this effect has been reported \cite{ahrens2025observation}.
A free-falling magnet also provides an opportunity to probe tidal effects induced by gravitational waves on spin systems \cite{MagneticWeberBar} in a genuine transverse–traceless (TT) gauge, and to experimentally examine the corresponding geodesic deviation predicted in the TT framework. Achieving such conditions presents two major challenges. First, creating and sustaining a microgravity environment is far more demanding than a tabletop setup. Second, the system must remain stably confined in the presence of residual random accelerations and noise introduced by free-fall platforms such as drop towers, sounding rockets, and space stations.

 Magnetic traps based on a PID-controlled coil were previously used to levitate magnets on Earth\,\cite{yadav2016optimized,el2002modeling,saitoh2020levitated,morikawa2004plasma}. However, such a configuration cannot operate under free-fall conditions, as we discuss here. In this work, we design and demonstrate a composite magnetic trap, a master-PID magnetic trap (MPIDMT) combining a PID-controlled coil system with a master control coil system. The system operates reliably under accelerations ranging from microgravity (free fall) to 2.5\,g (effectively, simulated in the lab) and can be finely tuned to compensate for acceleration shocks as high as 1.5\,g, with the potential for even higher stability in future experiments.

We carried out the first free-fall tests of this system using a third-generation drop-tower facility, the Einstein-Elevator in Hannover\,\cite{Lotz2022}.
During the microgravity phase, the magnet was successfully trapped and remained stable in a weak filed configuration (effectively 0.4\,g). Unlike mechanical methods, which may introduce electrostatic effects when releasing the magnet, the MPIDMT approach enables the magnet to enter pure free fall by removing the trapping fields without inducing a noticeable transient impulse. These results demonstrate the robustness of the trapping scheme and confirm that it operates reliably under both free-fall and high-acceleration–noise conditions. This achievement represents an important milestone toward next-generation magnetic sensing and paves the way for future demonstrations in space-based platforms.
 
\textit{PID controlled magnetic trap --} 

A PID-based magnetic trap is a widely used approach for achieving magnetic levitation\,\cite{yadav2016optimized,el2002modeling,saitoh2020levitated,morikawa2004plasma}. Compared with passive schemes, active PID control enables much faster stabilization of the magnet and offers precise, flexible tuning of key trapping parameters, including the resonance frequency, damping rate, and trap-center position. Such rapid and programmable stabilization is especially crucial in free-fall experiments, for instance, in the drop tower, where the system must endure a 5\,g acceleration and braking phase for each 0.5\,s.

\begin{figure}[htb]
    \centering
\includegraphics[width=0.99\linewidth]{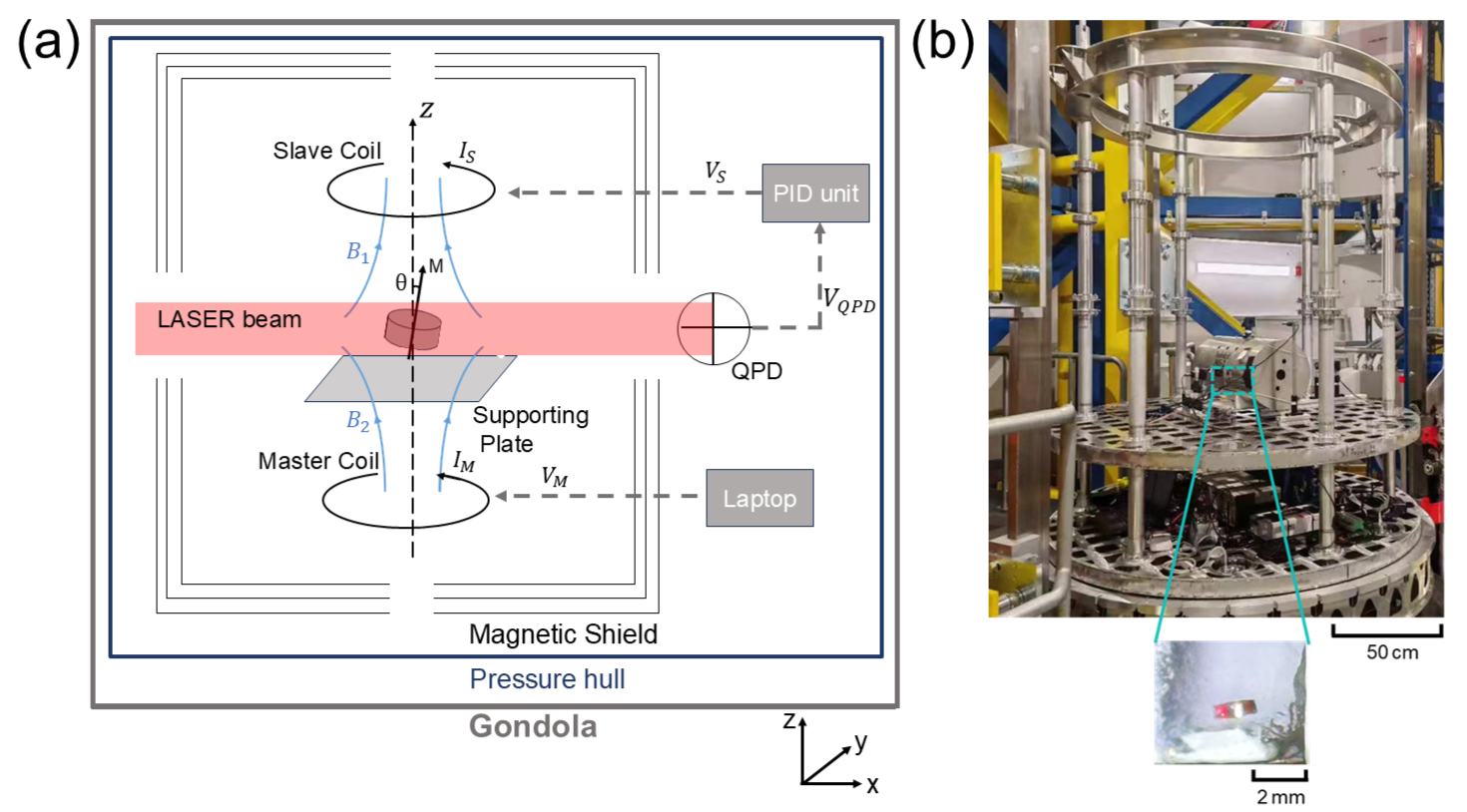}
\caption{Experimental setup in the Einstein-Elevator .
(a) Schematic of the setup with the master-PID magnetic trap (MPIDMT). A magnet is levitated between an upper slave coil and a lower master coil. The quadrant photodetector (QPD) is for position feedback.
(b) Top: a photograph of the experimental setup inside the Einstein-Elevator. The optical and PID control units and the silver-colored cylindrical magnetic shield are mounted on the carrier and are shown at the center. Bottom: side view of the the levitated magnet illuminated with a laser beam. 
}
\label{Fig.setup} 
\end{figure}

We used a conventional PID-controlled magnetic trap illustrated in Fig.\,\ref{Fig.setup} (with an additional lower master coil, which is specific to our composite-trap configuration). The upper PID-controlled (slave) coil is used to generate a vertical magnetic field to orientate the magnetic moment of the levitated magnet, while its field gradient exerts an upward magnetic force that balances gravity. A laser beam with a diameter larger than the magnet is directed onto the magnet so that the transmitted portion reaches a position-sensitive quadrant photodetector (QPD). The magnet casts a shadow on the transmitted beam, and any displacement of the magnet results in a corresponding change in the light intensity pattern on the QPD. The resulting voltage signal is therefore sensitive to the position of the magnet.

The PID controller processes the QPD signal and provides feedback to the current source driving the slave coil, thereby stabilizing the vertical movement of the magnet.  Because the acceleration along $z$ is proportional to the coil current, and the voltage of QPD $V_{\rm QPD}$ is linear with $z$ within the small-displacement regime (see SM \cite{SM}), the closed-loop dynamics can be approximately expressed as
\begin{equation}
\ddot{\text{z}}=-fP(\text{z}+D\dot{\text{z}})\,,
\label{Eq1} 
\end{equation}
where the integral term $I$ of the common PID is not used, the proportional ($P$) and derivative ($D$) terms, denoting the position and velocity feedback respectively, are retained. The constant $f$ absorbs the proportionality factors between the QPD signal, coil current, and the resulting magnetic force (see SM \cite{SM}). This expression represents a first-order linearized model. 
Deviations from linear behavior arise when the magnet moves slightly away from the expected center such spatial inhomogeneities of the magnetic-field geometry are visible or the magnet approaches the boundary of the linear detection range of the QPD.

To analyze the dynamics transverse to the $z$-axis, we use a first order expansion of the Lagrangian to express the translational motion
of the magnet along $x$ and its rotational motion $\theta$ within the $x-z$ plane. 
Translational stability along the $y$ direction, as well as rotations within the $y-z$ plane, can be treated analogously due to symmetry. For this first-order analysis, we neglect the coupling between these modes,
\begin{equation}
\begin{pmatrix} \ddot{x} \\ \ddot{\theta} \end{pmatrix}
= -
\begin{pmatrix} -\frac{M}{m}\partial_{\text{x}}^2B_{\text{z}} & -\frac{M}{m}\partial_{\text{x}}B_{\text{x}} \\ -\frac{M}{J}\partial_{\text{x}}B_{\text{x}} & \frac{MB_{\text{z}}}{J} \end{pmatrix}
\begin{pmatrix}x \\ \theta \end{pmatrix},\label{eq.stable.matrix}
\end{equation}
where $\partial_x B_z$ and $\partial_x B_x$ are the gradients of $B_z$ and $B_x$ along the $x$-axis, respectively. $M$, $m$, and $J$ denote the magnetic moment, mass, and moment of inertia of the levitated magnet. 
In order for the system to be stable in terms of translational and rotational degrees of freedom, the determinant and the trace must be positive, so that the matrix has two positive eigenvalues:
\begin{equation}
-\frac{M}{m}\partial_{\text{x}}^2B_{\text{z}}+\frac{MB_{\text{z}}}{J}>0\,,\label{eq.dia.positive}
\end{equation}
\begin{equation}
-B_{\text{z}}\cdot\partial_{\text{x}}^2B_{\text{z}}- (\partial_{\text{x}}B_{\text{x}})^2>0\,.\label{eq.det.positive}
\end{equation}

The slave coil generates a bias field $B_z$ that aligns the magnetic moment M and provides radial stabilization through the field curvature $-\partial_x^2B_z$\,\cite{simon2001diamagnetically}. On the ground, it must also supply a vertical gradient to counteract gravity, satisfying $M\,\partial_z B_z = mg$ at the trapping position. For the coil geometry used here (Fig.~\ref{Fig.setup}), -$\partial_x^2B_z$, $B_z$, and $\partial_zB_z$ share the same sign.
In this configuration, one finds that the left-hand side of inequality \eqref{eq.det.positive} is always positive, determined solely by the coil geometry (see SM \cite{SM}).

Inequality \eqref{eq.dia.positive}, however, depends on the sign of the magnetic-field derivatives. Under Earth gravity, we have $\partial_zB_z>0$ in most cases, with gravity acting as a bias to maintain the correct sign and ensure stability.
When the effective bias is reduced or reversed — for example, by disturbances exceeding the gravitational acceleration — $\partial_zB_z$ changes sign and both terms on the left-hand side of inequality\,\eqref{eq.dia.positive} also change sign
, and the inequality will no longer be satisfied. This issue becomes particularly severe in low-gravity or free-fall conditions, where the absence of a gravitational bias renders the trap highly sensitive to external perturbations.

It is therefore crucial to recognize that, without the gravity-provided bias acceleration, a conventional PID-controlled levitation trap becomes fundamentally unstable. To overcome this limitation, we introduce a coaxial bias coil beneath the magnet that generates a static bias field and sets the allowable range of the PID-controlled current, see inequality \eqref{eq.dia.positive}. This ensures stable trapping without reversing the magnetic field or its gradient. Since the PID current continuously adjusts in response to the bias-coil field, we refer to the bias coil as the master coil and the PID-controlled coil as the slave coil. 

\begin{figure}[htb]
    \centering
\includegraphics[width=0.99\linewidth]{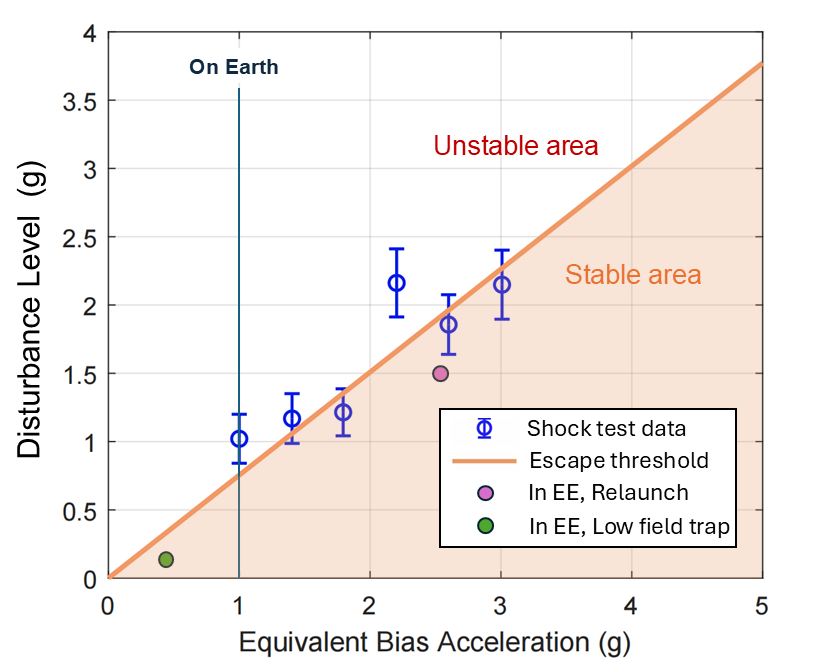}
\caption{Stability analysis under different equivalent bias accelerations. Both the applied disturbance and the bias field are expressed in units of $\rm g \approx 9.8\,\mathrm{m/s^2}$. When the shock-acceleration amplitude remains below the threshold—i.e., within the shaded region—the magnet stays stably trapped.}
\label{Fig.max acc vs current} 
\end{figure}

The stability of the system is determined by the reversal threshold of the PID-controlled coil—referred known as the escape threshold—at which the particle is no longer confined. This threshold is determined by the applied bias acceleration as shown in Fig.~\ref{Fig.max acc vs current}. To evaluate the stability of the trapping configuration, systematic shock tests were performed to determine the tolerance to external disturbances. The horizontal axis represents the effective bias acceleration, defined as 1g plus the downward acceleration generated by the master coil under laboratory conditions, and solely by the master coil in microgravity. For each selected bias, the master-coil current was incrementally adjusted, and shock tests were conducted to quantify the maximum disturbance the trap could withstand. This procedure yields the stability limit as a function of bias. In the optimal configuration, the trap is expected to remain stable under disturbances up to approximately 3.5\,g at a bias of 5\,g.

Because the master coil current can be adjusted, the bias can be dynamically tuned—set small under gentle conditions such as free-fall stages with minor residual accelerations, and increased during high-disturbance periods such as the relaunch phase in a drop-tower experiment.

\begin{figure}[htb]
    \centering
\includegraphics[width=0.99\linewidth]{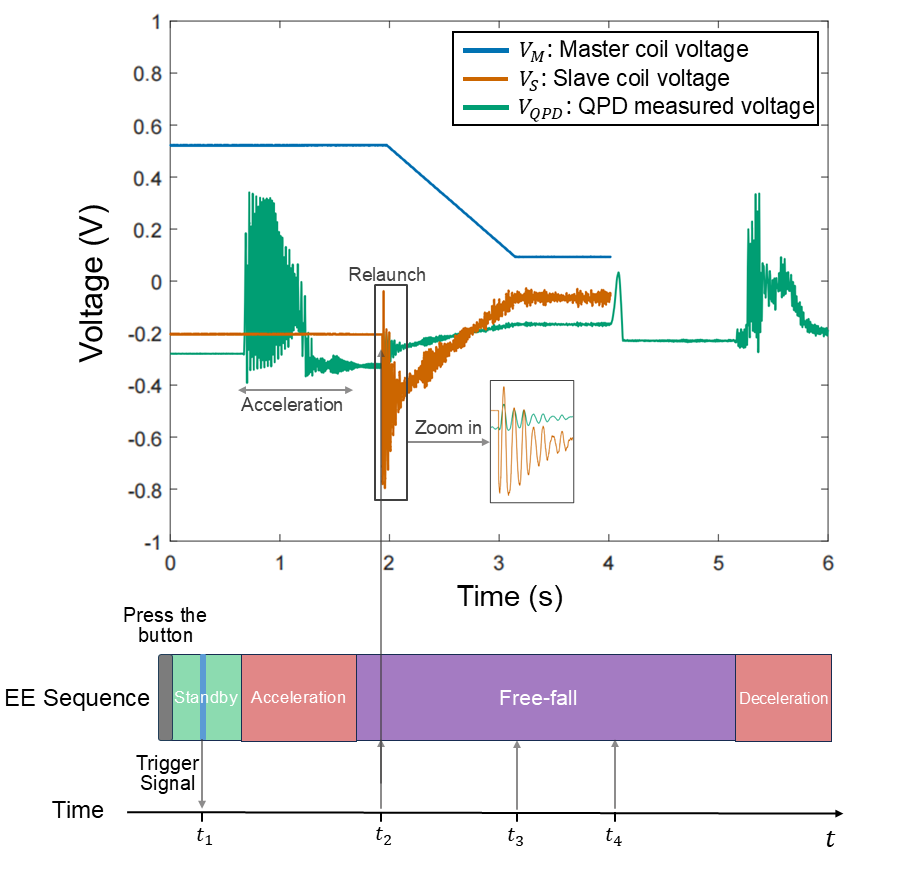}
\caption{Programmed sequence control on MPIDMT and the sequence from Einstein-Elevator to be synchronized. The lower panel shows the programmed profile of the EE, including the launch (acceleration), free fall, and deceleration phases. The upper trace displays the voltages for the master (blue) and slave (orange) coils, alongside the QPD measured voltage (green), marking the critical events: receipt of the trigger signal at time $t_1$, activation of PID control at $t_2$, reduction of the bias field at $t_3$, and cutoff of all currents at $t_4$. The QPD-signal fluctuations during the EE launch and deceleration result from vibration-induced noise in the optical system. During $t_3$ to $t_4$ in microgravity phase, the stabilization of both the slave coil voltage and the QPD signal evidences a successful capture and stable confinement of the relaunched magnet by the MPIDMT trap. The peak after $t_4$ arises from the QPD measurement system (see SM \cite{SM}), indicating that the magnet moves upward until it is out of the detection range.}
\label{Fig.sequence} 
\end{figure} 

 \textit{ Free Fall Experiment--}

 The operation sequence of the Einstein-Elevator (EE) is illustrated in the lower panel of Fig.~\ref{Fig.sequence}. Each cycle begins with a trigger signal from the control room at time $t_1$, initiating an acceleration phase (5g) of 0.5\,s during which the EE is brought up to its launch velocity. This is followed by a free-fall interval of about 4\,s, providing a microgravity environment inside the cabin. During the free-fall phase, the EE continues its parabolic trajectory, moving upward to its apex and then descending back down under gravity. It then enters a final deceleration stage, which brings the system smoothly back to its initial position at rest. Acceleration profiles of different flights are shown in the SM \cite{SM}, where the free-falling quality highly depends on the vacuum between pressure hull and the gondola in each flight.

To synchronize magnetic levitation with the free-fall sequence of the Einstein-Elevator, we implemented a programmable control routine. Upon receiving the trigger signal from the EE, the control program initiates a predefined timing sequence based on the system clock. During the EE acceleration phase, stable levitation is not feasible due to the large perturbations, up to 5\,g, during which pronounced voltage fluctuations are observed, originating from vibrations of the optical components even in the absence of a levitated magnet; therefore, instead of attempting active feedback, we hold the magnet on the supporting plate with a static bias field throughout this phase. 

Approximately 0.1\,s after the acceleration ends—at time $t_2$—the controller switches from manual mode to PID control. This handover produces a brief upward impulse on the magnet, after which the PID trap, configured with high proportional gain and an initially large master-coil current, rapidly recaptures and stabilizes the magnet within about 0.2\,s. The results are shown in Fig.\,\ref{Fig.sequence} and labeled as relaunch. In contrast to the launch of the Einstein-Elevator, the relaunch refers to the launch of the levitated magnet itself. With a bias field corresponding to an effective acceleration of approximately 2.5\,g, the trap successfully stabilized the magnet against a pulsed acceleration disturbance of about 1.5\,g.

Once the magnet is trapped, we gradually reduce the master-coil current, lowering the effective bias acceleration and allowing the magnet to transition into the low-field trap  configuration at time $t_3$ with bias acceleration of approximately 0.4\,g. 

\begin{figure}[htbp]
  \centering  {\includegraphics[width=0.5\textwidth]{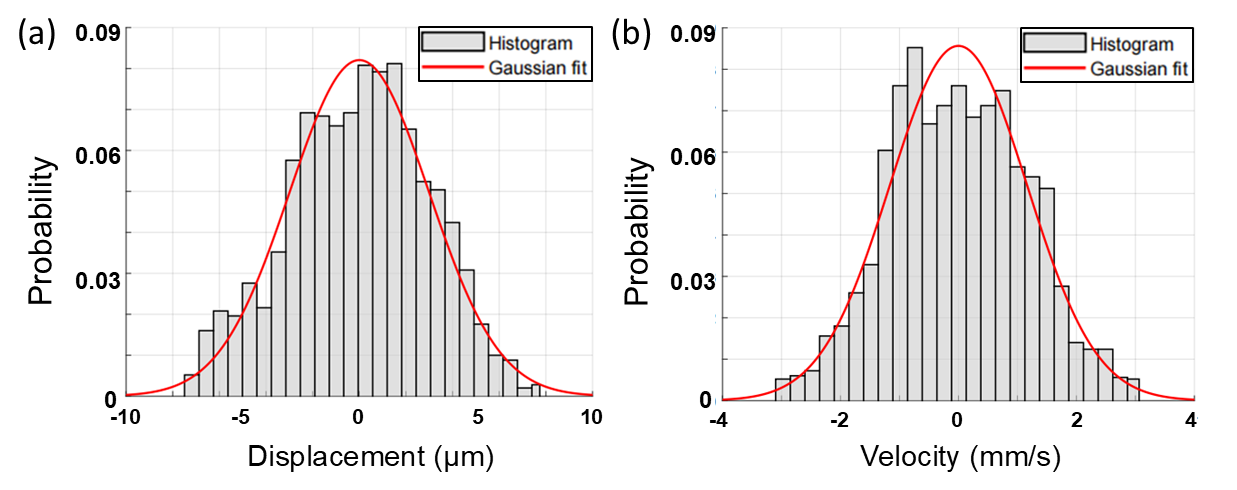}}
  \caption{Probability distributions of the (a) position and (b) velocity of the magnet during the trapped intervals $t_3$ and $t_4$. Both distributions are well described by Gaussian functions. The 90\% confidence intervals correspond to spatial and velocity spreads of 4.9\,$\mu\text{m}$ and 1.9\,$\text{mm/s}$, respectively.}
\label{fig.distribution.velocity.position}
\end{figure}

The noise during free-fall trapping can be modeled as additional terms in Eq.\,\eqref{Eq1}:
\begin{equation}
\ddot{z}+(\gamma+fPD)\dot{z}+fPz=fP(D\dot{z_N}+z_N)+a_{\text{res}}(EE)\,,
\label{EQ5}
\end{equation}
where the right-hand side of the equation elucidates the primary noise sources and their coupling mechanisms within the system, encompassing both the optical vibration noise of the laser, $z_{N}$,  amplified via the feedback loop and the residual acceleration, $a_{\text{res}}(EE)$, from the Einstein-Elevator. $\gamma$ is defined as the damping coefficient, dominated by the viscous damping on the magnet due to its motion in the ambient air. 

\begin{figure}[htb]
    \centering
\includegraphics[width=0.99\linewidth]{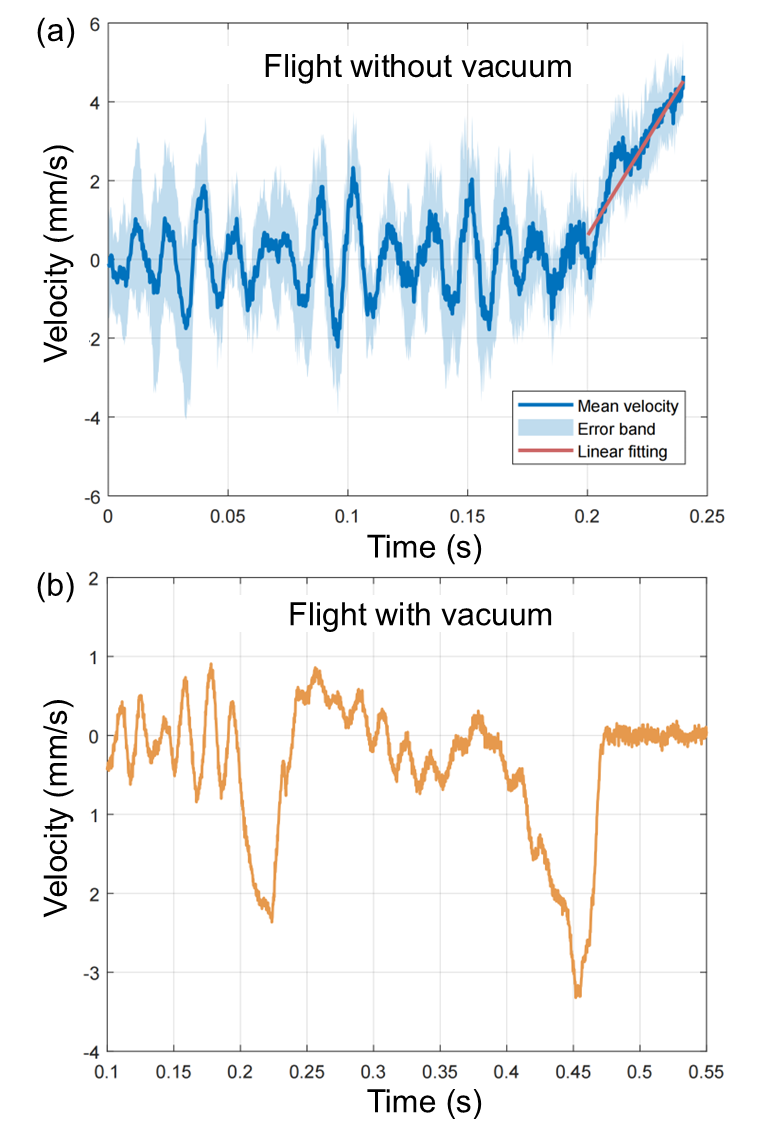}
\caption{Velocity profile| (a) Plot obtained by averaging different flights with ambient pressure in the gondola. The current cut off happens at $t_0=0.2 \, \text{s}  ~(t_4$ in Fig.\,\ref{Fig.sequence}), a mean residual velocity of $0.6\,\text{mm}/\text{s}$ at the cut off moment mainly comes from oscillation inside the magnetic trap. A clear residual acceleration upward is observed, with a mean value of $0.098\,\text{m}/\text{s}^2$. (b) Velocity profile after releasing the magnet for flight under a vacuum of 10\,mbar between pressure hull and the gondola. The current is switched off at the same moment, $t_0 = 0.2\,\text{s}$ ($t_4$ in Fig.\,\ref{Fig.sequence}). }
\label{Fig.velocity.change} 
\end{figure}

The spatial distribution of the magnet position and its corresponding velocity distribution during low-field trapping between $t_3$ and $t_4$ is presented in Fig.\,\ref{fig.distribution.velocity.position}. The trap confines the magnet within a position spread of $4.9\,\mu$m, while maintaining its velocity within $1.9\,$mm/s (90\%\,C.L. from a Gaussian distribution). The results demonstrate the ability of the MPIDMT architecture to stably trap the magnet at a low bias field of 0.4\,g. The dominant noise limiting the trapping stability originates from the laser-beam instabilities $z_N$, which is then amplified through the feedback loop, as shown in Eq.\,\ref{EQ5}. After maintaining stable levitation for an additional 0.8\,s, all coil currents are cut off at time $t_4$ via a relay switch. Figure\,\ref{Fig.velocity.change}\,(a) shows the reconstructed velocity profile of the magnet during trapping and release, obtained from four successful normal free-fall runs. Owing to the limited number of trials, residual oscillatory features remain visible in the velocity plot, arising from the resonant response of the PID trap and from optical vibration noise near its mechanical resonance.

Ideally, the magnet would remain freely floating for a longer time within the trap region after the current is cut-off. However, the measured velocity after $t_4$ reveals a residual acceleration, likely due to aerodynamic coupling between the gondola and the pressure hull, as the system is not under vacuum. The mean acceleration after cut-off is approximately 0.01\,g, consistent with the residual acceleration level of $10^{-2}\,\mathrm{g}$ from the directly measured motion of the Einstein-Elevator. (see SM \cite{SM}) This acceleration causes the magnet to drift out of the optical detection range after about 0.04\,s, which is also shown by the position peak in the QPD signal shown in Fig.\,\ref{Fig.sequence}. 

To confirm this suspicion, we performed additional flights under vacuum in the EE. A larger dataset from vacuum flights would be necessary for a statistically robust analysis. However, due to time constraints, only a limited number of tests could be performed. A representative successful dataset is shown in Fig.\,\ref{Fig.velocity.change}\,(b), demonstrating a free-floating duration of approximately 0.3\,s. In contrast to the non-vacuum flights, the persistent upward residual acceleration is no longer observed under vacuum conditions. Note that the vacuum was applied only to the space between the pressure hull and the gondola, while the experimental apparatus inside the pressure hull remained at ambient pressure throughout all flights. These observations indicate that the quality of free fall in the EE remains a major limitation to achieving a truly free-floating magnet. In the meantime, a residual velocity may also be introduced when the trapping currents are switched off. This may arise from oscillations within the trap, as well as transient effects associated with the current switch-off. From the statistical fit shown in Fig.\,\ref{Fig.velocity.change}\,(a), together with the measured velocity at the release moment in Fig.\,\ref{Fig.velocity.change}\,(b), the residual velocity is found to remain below $1.9\,\text{mm}/\text{s}$ at the 90\% confidence level from the oscillatory distribution. Given the limited statistics, it is not possible to quantitatively isolate the contribution from transient effects associated with the current switch-off. However, an upper-bound estimate suggests that this contribution remains below $1.7\,\text{mm}/\text{s}$.

Possible causes of transient impact on the magnet at cutting off moment are time deviations between the cut-off times of the two coils and transient electromagnetic effects at the moment of release. For both, cutting off at low trapping field reduces the effect compared to a large bias field. The minimum attainable trapping field is ultimately limited by the noise level of the PID control system. In our current setup, the lowest stably maintainable equivalent field is approximately 0.4\,g.

For the PID trap parameters, a larger proportional gain $P$ is essential for faster trapping and better confinement, since it increases the trap frequency and shortens the stabilization time. However, the range of the slave coil current determines the upper limit of $P$. During the kick-up stage, an unavoidable position shift occurs; if $P$ is set too high, the PID coil can overcompensate and reverse the magnetic field, even in the presence of the master coil. Thus, achieving stable trapping at the lowest bias field requires careful optimization of the bias field and P gain. The experimentally optimized control parameters are provided in the SM \cite{SM}.

\textit{Conclusion}| In conclusion, we have demonstrated robust feedback levitation of a magnet in the free-fall environment of the Einstein-Elevator using a novel MPIDMT. This architecture successfully resolves the inherent instability of conventional traps in microgravity, allowing stable operation against shocks up to 1.5\,g and precise tuning to a low-field (0.4\,g) configuration. The capability to actively control the trap stiffness and release the magnet into a near-true free-fall state is a key enabler for future high-sensitivity experiments. While the current drop-tower platform provides a crucial proof-of-principle, its limited near-zero gravity duration (4\,s) presents a constraint that can be naturally overcome by implementing a similar setup on a space-based platform like a space station. Future engineering efforts must focus on miniaturizing the electronics and optimizing for power consumption, leveraging advances from space-borne atomic and cold-atom experiments  \cite{2018In,cacciapuoti2024atomic,aveline2020observation,elliott2018nasa,li2025realization,he2023space}. Deployed in space with pure free fall environment, a levitated ferromagnetic sensor would achieve unprecedented sensitivity, finally bringing the observation of pure macroscopic Larmor precession within reach. This capability would revolutionize our ability to map ambient magnetic fields and monitor onboard electronics activity, while also serving as a pathfinder for fundamental inquiries—from the search for dark matter to testing general relativity's effects on quantum spins \cite{fadeev2021gravity} in a pristine inertial frame.

\section*{Acknowledgment}
The authors acknowledge the support of this project by Prof. E.\,Rasel. 
This work was supported by the QuantERA project
LEMAQUME (DFG Project No. 500314265), as well as by the
 Cluster of Excellence ``Precision Physics, Fundamental Interactions, and Structure of Matter'' (PRISMA++ EXC 2118/2) funded by the German Research Foundation (DFG) within the German Excellence Strategy (Project ID 390831469), and by the COST Action within the project COSMIC WISPers (Grant No. CA21106). W. J. acknowledges the support from The Fundamental Research Funds for the Central Universities, Peking University. Y. L. acknowledges financial support from the China Scholarship Council (CSC) program.

\bibliographystyle{apsrev4-1}

\begin{thebibliography}{36}%
\makeatletter
\providecommand \@ifxundefined [1]{%
 \@ifx{#1\undefined}
}%
\providecommand \@ifnum [1]{%
 \ifnum #1\expandafter \@firstoftwo
 \else \expandafter \@secondoftwo
 \fi
}%
\providecommand \@ifx [1]{%
 \ifx #1\expandafter \@firstoftwo
 \else \expandafter \@secondoftwo
 \fi
}%
\providecommand \natexlab [1]{#1}%
\providecommand \enquote  [1]{``#1''}%
\providecommand \bibnamefont  [1]{#1}%
\providecommand \bibfnamefont [1]{#1}%
\providecommand \citenamefont [1]{#1}%
\providecommand \href@noop [0]{\@secondoftwo}%
\providecommand \href [0]{\begingroup \@sanitize@url \@href}%
\providecommand \@href[1]{\@@startlink{#1}\@@href}%
\providecommand \@@href[1]{\endgroup#1\@@endlink}%
\providecommand \@sanitize@url [0]{\catcode `\\12\catcode `\$12\catcode
  `\&12\catcode `\#12\catcode `\^12\catcode `\_12\catcode `\%12\relax}%
\providecommand \@@startlink[1]{}%
\providecommand \@@endlink[0]{}%
\providecommand \url  [0]{\begingroup\@sanitize@url \@url }%
\providecommand \@url [1]{\endgroup\@href {#1}{\urlprefix }}%
\providecommand \urlprefix  [0]{URL }%
\providecommand \Eprint [0]{\href }%
\providecommand \doibase [0]{http://dx.doi.org/}%
\providecommand \selectlanguage [0]{\@gobble}%
\providecommand \bibinfo  [0]{\@secondoftwo}%
\providecommand \bibfield  [0]{\@secondoftwo}%
\providecommand \translation [1]{[#1]}%
\providecommand \BibitemOpen [0]{}%
\providecommand \bibitemStop [0]{}%
\providecommand \bibitemNoStop [0]{.\EOS\space}%
\providecommand \EOS [0]{\spacefactor3000\relax}%
\providecommand \BibitemShut  [1]{\csname bibitem#1\endcsname}%
\let\auto@bib@innerbib\@empty
\bibitem [{\citenamefont {Boto}\ \emph {et~al.}(2018)\citenamefont {Boto},
  \citenamefont {Holmes}, \citenamefont {Leggett}, \citenamefont {Roberts},
  \citenamefont {Shah}, \citenamefont {Meyer}, \citenamefont {Mu{\~n}oz},
  \citenamefont {Mullinger}, \citenamefont {Tierney}, \citenamefont {Bestmann}
  \emph {et~al.}}]{boto2018moving}%
  \BibitemOpen
  \bibfield  {author} {\bibinfo {author} {\bibfnamefont {E.}~\bibnamefont
  {Boto}}, \bibinfo {author} {\bibfnamefont {N.}~\bibnamefont {Holmes}},
  \bibinfo {author} {\bibfnamefont {J.}~\bibnamefont {Leggett}}, \bibinfo
  {author} {\bibfnamefont {G.}~\bibnamefont {Roberts}}, \bibinfo {author}
  {\bibfnamefont {V.}~\bibnamefont {Shah}}, \bibinfo {author} {\bibfnamefont
  {S.~S.}\ \bibnamefont {Meyer}}, \bibinfo {author} {\bibfnamefont {L.~D.}\
  \bibnamefont {Mu{\~n}oz}}, \bibinfo {author} {\bibfnamefont {K.~J.}\
  \bibnamefont {Mullinger}}, \bibinfo {author} {\bibfnamefont {T.~M.}\
  \bibnamefont {Tierney}}, \bibinfo {author} {\bibfnamefont {S.}~\bibnamefont
  {Bestmann}},  \emph {et~al.},\ }\href@noop {} {\bibfield  {journal} {\bibinfo
   {journal} {Nature}\ }\textbf {\bibinfo {volume} {555}},\ \bibinfo {pages}
  {657} (\bibinfo {year} {2018})}\BibitemShut {NoStop}%
\bibitem [{\citenamefont {Cohen}(1972)}]{cohen1972magnetoencephalography}%
  \BibitemOpen
  \bibfield  {author} {\bibinfo {author} {\bibfnamefont {D.}~\bibnamefont
  {Cohen}},\ }\href@noop {} {\bibfield  {journal} {\bibinfo  {journal}
  {Science}\ }\textbf {\bibinfo {volume} {175}},\ \bibinfo {pages} {664}
  (\bibinfo {year} {1972})}\BibitemShut {NoStop}%
\bibitem [{\citenamefont {Aslam}\ \emph {et~al.}(2023)\citenamefont {Aslam},
  \citenamefont {Zhou}, \citenamefont {Urbach}, \citenamefont {Turner},
  \citenamefont {Walsworth}, \citenamefont {Lukin},\ and\ \citenamefont
  {Park}}]{aslam2023quantum}%
  \BibitemOpen
  \bibfield  {author} {\bibinfo {author} {\bibfnamefont {N.}~\bibnamefont
  {Aslam}}, \bibinfo {author} {\bibfnamefont {H.}~\bibnamefont {Zhou}},
  \bibinfo {author} {\bibfnamefont {E.~K.}\ \bibnamefont {Urbach}}, \bibinfo
  {author} {\bibfnamefont {M.~J.}\ \bibnamefont {Turner}}, \bibinfo {author}
  {\bibfnamefont {R.~L.}\ \bibnamefont {Walsworth}}, \bibinfo {author}
  {\bibfnamefont {M.~D.}\ \bibnamefont {Lukin}}, \ and\ \bibinfo {author}
  {\bibfnamefont {H.}~\bibnamefont {Park}},\ }\href@noop {} {\bibfield
  {journal} {\bibinfo  {journal} {Nature Reviews Physics}\ }\textbf {\bibinfo
  {volume} {5}},\ \bibinfo {pages} {157} (\bibinfo {year} {2023})}\BibitemShut
  {NoStop}%
\bibitem [{\citenamefont {Wei}\ \emph {et~al.}(2023)\citenamefont {Wei},
  \citenamefont {Zhao}, \citenamefont {Fang}, \citenamefont {Xu}, \citenamefont
  {Liu}, \citenamefont {Cao}, \citenamefont {Wickenbrock}, \citenamefont {Hu},
  \citenamefont {Ji}, \citenamefont {Fang} \emph
  {et~al.}}]{wei2023ultrasensitive}%
  \BibitemOpen
  \bibfield  {author} {\bibinfo {author} {\bibfnamefont {K.}~\bibnamefont
  {Wei}}, \bibinfo {author} {\bibfnamefont {T.}~\bibnamefont {Zhao}}, \bibinfo
  {author} {\bibfnamefont {X.}~\bibnamefont {Fang}}, \bibinfo {author}
  {\bibfnamefont {Z.}~\bibnamefont {Xu}}, \bibinfo {author} {\bibfnamefont
  {C.}~\bibnamefont {Liu}}, \bibinfo {author} {\bibfnamefont {Q.}~\bibnamefont
  {Cao}}, \bibinfo {author} {\bibfnamefont {A.}~\bibnamefont {Wickenbrock}},
  \bibinfo {author} {\bibfnamefont {Y.}~\bibnamefont {Hu}}, \bibinfo {author}
  {\bibfnamefont {W.}~\bibnamefont {Ji}}, \bibinfo {author} {\bibfnamefont
  {J.}~\bibnamefont {Fang}},  \emph {et~al.},\ }\href@noop {} {\bibfield
  {journal} {\bibinfo  {journal} {Physical Review Letters}\ }\textbf {\bibinfo
  {volume} {130}},\ \bibinfo {pages} {063201} (\bibinfo {year}
  {2023})}\BibitemShut {NoStop}%
\bibitem [{\citenamefont {Xu}\ \emph {et~al.}(2024)\citenamefont {Xu},
  \citenamefont {Ma}, \citenamefont {Wei}, \citenamefont {He}, \citenamefont
  {Heng}, \citenamefont {Huang}, \citenamefont {Ai}, \citenamefont {Liao},
  \citenamefont {Ji}, \citenamefont {Liu} \emph {et~al.}}]{xu2024constraining}%
  \BibitemOpen
  \bibfield  {author} {\bibinfo {author} {\bibfnamefont {Z.}~\bibnamefont
  {Xu}}, \bibinfo {author} {\bibfnamefont {X.}~\bibnamefont {Ma}}, \bibinfo
  {author} {\bibfnamefont {K.}~\bibnamefont {Wei}}, \bibinfo {author}
  {\bibfnamefont {Y.}~\bibnamefont {He}}, \bibinfo {author} {\bibfnamefont
  {X.}~\bibnamefont {Heng}}, \bibinfo {author} {\bibfnamefont {X.}~\bibnamefont
  {Huang}}, \bibinfo {author} {\bibfnamefont {T.}~\bibnamefont {Ai}}, \bibinfo
  {author} {\bibfnamefont {J.}~\bibnamefont {Liao}}, \bibinfo {author}
  {\bibfnamefont {W.}~\bibnamefont {Ji}}, \bibinfo {author} {\bibfnamefont
  {J.}~\bibnamefont {Liu}},  \emph {et~al.},\ }\href@noop {} {\bibfield
  {journal} {\bibinfo  {journal} {Communications Physics}\ }\textbf {\bibinfo
  {volume} {7}},\ \bibinfo {pages} {226} (\bibinfo {year} {2024})}\BibitemShut
  {NoStop}%
\bibitem [{\citenamefont {Ji}\ \emph {et~al.}(2018)\citenamefont {Ji},
  \citenamefont {Chen}, \citenamefont {Fu}, \citenamefont {Ding}, \citenamefont
  {Fang}, \citenamefont {Xiao}, \citenamefont {Wei},\ and\ \citenamefont
  {Yan}}]{ji2018new}%
  \BibitemOpen
  \bibfield  {author} {\bibinfo {author} {\bibfnamefont {W.}~\bibnamefont
  {Ji}}, \bibinfo {author} {\bibfnamefont {Y.}~\bibnamefont {Chen}}, \bibinfo
  {author} {\bibfnamefont {C.}~\bibnamefont {Fu}}, \bibinfo {author}
  {\bibfnamefont {M.}~\bibnamefont {Ding}}, \bibinfo {author} {\bibfnamefont
  {J.}~\bibnamefont {Fang}}, \bibinfo {author} {\bibfnamefont {Z.}~\bibnamefont
  {Xiao}}, \bibinfo {author} {\bibfnamefont {K.}~\bibnamefont {Wei}}, \ and\
  \bibinfo {author} {\bibfnamefont {H.}~\bibnamefont {Yan}},\ }\href@noop {}
  {\bibfield  {journal} {\bibinfo  {journal} {Physical review letters}\
  }\textbf {\bibinfo {volume} {121}},\ \bibinfo {pages} {261803} (\bibinfo
  {year} {2018})}\BibitemShut {NoStop}%
\bibitem [{\citenamefont {Cong}\ \emph {et~al.}(2025)\citenamefont {Cong},
  \citenamefont {Ji}, \citenamefont {Fadeev}, \citenamefont {Ficek},
  \citenamefont {Jiang}, \citenamefont {Flambaum}, \citenamefont {Guan},
  \citenamefont {Jackson~Kimball}, \citenamefont {Kozlov}, \citenamefont
  {Stadnik} \emph {et~al.}}]{cong2025spin}%
  \BibitemOpen
  \bibfield  {author} {\bibinfo {author} {\bibfnamefont {L.}~\bibnamefont
  {Cong}}, \bibinfo {author} {\bibfnamefont {W.}~\bibnamefont {Ji}}, \bibinfo
  {author} {\bibfnamefont {P.}~\bibnamefont {Fadeev}}, \bibinfo {author}
  {\bibfnamefont {F.}~\bibnamefont {Ficek}}, \bibinfo {author} {\bibfnamefont
  {M.}~\bibnamefont {Jiang}}, \bibinfo {author} {\bibfnamefont {V.~V.}\
  \bibnamefont {Flambaum}}, \bibinfo {author} {\bibfnamefont {H.}~\bibnamefont
  {Guan}}, \bibinfo {author} {\bibfnamefont {D.~F.}\ \bibnamefont
  {Jackson~Kimball}}, \bibinfo {author} {\bibfnamefont {M.~G.}\ \bibnamefont
  {Kozlov}}, \bibinfo {author} {\bibfnamefont {Y.~V.}\ \bibnamefont {Stadnik}},
   \emph {et~al.},\ }\href@noop {} {\bibfield  {journal} {\bibinfo  {journal}
  {Reviews of Modern Physics}\ }\textbf {\bibinfo {volume} {97}},\ \bibinfo
  {pages} {025005} (\bibinfo {year} {2025})}\BibitemShut {NoStop}%
\bibitem [{\citenamefont {Kimball}\ \emph {et~al.}(2016)\citenamefont
  {Kimball}, \citenamefont {Sushkov},\ and\ \citenamefont
  {Budker}}]{kimball2016precessing}%
  \BibitemOpen
  \bibfield  {author} {\bibinfo {author} {\bibfnamefont {D.~F.~J.}\
  \bibnamefont {Kimball}}, \bibinfo {author} {\bibfnamefont {A.~O.}\
  \bibnamefont {Sushkov}}, \ and\ \bibinfo {author} {\bibfnamefont
  {D.}~\bibnamefont {Budker}},\ }\href@noop {} {\bibfield  {journal} {\bibinfo
  {journal} {Physical review letters}\ }\textbf {\bibinfo {volume} {116}},\
  \bibinfo {pages} {190801} (\bibinfo {year} {2016})}\BibitemShut {NoStop}%
\bibitem [{\citenamefont {Band}\ \emph {et~al.}(2018)\citenamefont {Band},
  \citenamefont {Avishai},\ and\ \citenamefont {Shnirman}}]{band2018dynamics}%
  \BibitemOpen
  \bibfield  {author} {\bibinfo {author} {\bibfnamefont {Y.}~\bibnamefont
  {Band}}, \bibinfo {author} {\bibfnamefont {Y.}~\bibnamefont {Avishai}}, \
  and\ \bibinfo {author} {\bibfnamefont {A.}~\bibnamefont {Shnirman}},\
  }\href@noop {} {\bibfield  {journal} {\bibinfo  {journal} {Physical Review
  Letters}\ }\textbf {\bibinfo {volume} {121}},\ \bibinfo {pages} {160801}
  (\bibinfo {year} {2018})}\BibitemShut {NoStop}%
\bibitem [{\citenamefont {Vinante}\ \emph {et~al.}(2021)\citenamefont
  {Vinante}, \citenamefont {Timberlake}, \citenamefont {Budker}, \citenamefont
  {Kimball}, \citenamefont {Sushkov},\ and\ \citenamefont
  {Ulbricht}}]{vinante2021surpassing}%
  \BibitemOpen
  \bibfield  {author} {\bibinfo {author} {\bibfnamefont {A.}~\bibnamefont
  {Vinante}}, \bibinfo {author} {\bibfnamefont {C.}~\bibnamefont {Timberlake}},
  \bibinfo {author} {\bibfnamefont {D.}~\bibnamefont {Budker}}, \bibinfo
  {author} {\bibfnamefont {D.~F.~J.}\ \bibnamefont {Kimball}}, \bibinfo
  {author} {\bibfnamefont {A.~O.}\ \bibnamefont {Sushkov}}, \ and\ \bibinfo
  {author} {\bibfnamefont {H.}~\bibnamefont {Ulbricht}},\ }\href@noop {}
  {\bibfield  {journal} {\bibinfo  {journal} {Physical Review Letters}\
  }\textbf {\bibinfo {volume} {127}},\ \bibinfo {pages} {070801} (\bibinfo
  {year} {2021})}\BibitemShut {NoStop}%
\bibitem [{\citenamefont {Ni}\ \emph {et~al.}(2025)\citenamefont {Ni},
  \citenamefont {Zou}, \citenamefont {Lecamwasam}, \citenamefont {Vinante},
  \citenamefont {Budker}, \citenamefont {Lam}, \citenamefont {Wang},\ and\
  \citenamefont {Gong}}]{ni2025microscopic}%
  \BibitemOpen
  \bibfield  {author} {\bibinfo {author} {\bibfnamefont {X.}~\bibnamefont
  {Ni}}, \bibinfo {author} {\bibfnamefont {Z.}~\bibnamefont {Zou}}, \bibinfo
  {author} {\bibfnamefont {R.}~\bibnamefont {Lecamwasam}}, \bibinfo {author}
  {\bibfnamefont {A.}~\bibnamefont {Vinante}}, \bibinfo {author} {\bibfnamefont
  {D.}~\bibnamefont {Budker}}, \bibinfo {author} {\bibfnamefont {P.~K.}\
  \bibnamefont {Lam}}, \bibinfo {author} {\bibfnamefont {T.}~\bibnamefont
  {Wang}}, \ and\ \bibinfo {author} {\bibfnamefont {J.}~\bibnamefont {Gong}},\
  }\href@noop {} {\bibfield  {journal} {\bibinfo  {journal} {Physical Review
  Research}\ }\textbf {\bibinfo {volume} {7}},\ \bibinfo {pages} {043120}
  (\bibinfo {year} {2025})}\BibitemShut {NoStop}%
\bibitem [{\citenamefont {Ji}\ \emph {et~al.}(2025)\citenamefont {Ji},
  \citenamefont {Xu}, \citenamefont {Qu},\ and\ \citenamefont
  {Budker}}]{ji2025levitated}%
  \BibitemOpen
  \bibfield  {author} {\bibinfo {author} {\bibfnamefont {W.}~\bibnamefont
  {Ji}}, \bibinfo {author} {\bibfnamefont {C.}~\bibnamefont {Xu}}, \bibinfo
  {author} {\bibfnamefont {G.}~\bibnamefont {Qu}}, \ and\ \bibinfo {author}
  {\bibfnamefont {D.}~\bibnamefont {Budker}},\ }\href@noop {} {\bibfield
  {journal} {\bibinfo  {journal} {arXiv preprint arXiv:2504.21524}\ } (\bibinfo
  {year} {2025})}\BibitemShut {NoStop}%
\bibitem [{\citenamefont {Allred}\ \emph {et~al.}(2002)\citenamefont {Allred},
  \citenamefont {Lyman}, \citenamefont {Kornack},\ and\ \citenamefont
  {Romalis}}]{allred2002high}%
  \BibitemOpen
  \bibfield  {author} {\bibinfo {author} {\bibfnamefont {J.}~\bibnamefont
  {Allred}}, \bibinfo {author} {\bibfnamefont {R.}~\bibnamefont {Lyman}},
  \bibinfo {author} {\bibfnamefont {T.}~\bibnamefont {Kornack}}, \ and\
  \bibinfo {author} {\bibfnamefont {M.~V.}\ \bibnamefont {Romalis}},\
  }\href@noop {} {\bibfield  {journal} {\bibinfo  {journal} {Physical review
  letters}\ }\textbf {\bibinfo {volume} {89}},\ \bibinfo {pages} {130801}
  (\bibinfo {year} {2002})}\BibitemShut {NoStop}%
\bibitem [{\citenamefont {Clarke}\ and\ \citenamefont
  {Braginski}(2006)}]{clarke2006squid}%
  \BibitemOpen
  \bibfield  {author} {\bibinfo {author} {\bibfnamefont {J.}~\bibnamefont
  {Clarke}}\ and\ \bibinfo {author} {\bibfnamefont {A.~I.}\ \bibnamefont
  {Braginski}},\ }\href@noop {} {\emph {\bibinfo {title} {The SQUID handbook:
  Applications of SQUIDs and SQUID systems}}}\ (\bibinfo  {publisher} {John
  Wiley \& Sons},\ \bibinfo {year} {2006})\BibitemShut {NoStop}%
\bibitem [{\citenamefont {Wang}\ \emph {et~al.}(2019)\citenamefont {Wang},
  \citenamefont {Lourette}, \citenamefont {O’Kelley}, \citenamefont {Kayci},
  \citenamefont {Band}, \citenamefont {Kimball}, \citenamefont {Sushkov},\ and\
  \citenamefont {Budker}}]{wang2019dynamics}%
  \BibitemOpen
  \bibfield  {author} {\bibinfo {author} {\bibfnamefont {T.}~\bibnamefont
  {Wang}}, \bibinfo {author} {\bibfnamefont {S.}~\bibnamefont {Lourette}},
  \bibinfo {author} {\bibfnamefont {S.~R.}\ \bibnamefont {O’Kelley}},
  \bibinfo {author} {\bibfnamefont {M.}~\bibnamefont {Kayci}}, \bibinfo
  {author} {\bibfnamefont {Y.}~\bibnamefont {Band}}, \bibinfo {author}
  {\bibfnamefont {D.~F.~J.}\ \bibnamefont {Kimball}}, \bibinfo {author}
  {\bibfnamefont {A.~O.}\ \bibnamefont {Sushkov}}, \ and\ \bibinfo {author}
  {\bibfnamefont {D.}~\bibnamefont {Budker}},\ }\href@noop {} {\bibfield
  {journal} {\bibinfo  {journal} {Physical Review Applied}\ }\textbf {\bibinfo
  {volume} {11}},\ \bibinfo {pages} {044041} (\bibinfo {year}
  {2019})}\BibitemShut {NoStop}%
\bibitem [{\citenamefont {Vinante}\ \emph {et~al.}(2020)\citenamefont
  {Vinante}, \citenamefont {Falferi}, \citenamefont {Gasbarri}, \citenamefont
  {Setter}, \citenamefont {Timberlake},\ and\ \citenamefont
  {Ulbricht}}]{vinante2020ultralow}%
  \BibitemOpen
  \bibfield  {author} {\bibinfo {author} {\bibfnamefont {A.}~\bibnamefont
  {Vinante}}, \bibinfo {author} {\bibfnamefont {P.}~\bibnamefont {Falferi}},
  \bibinfo {author} {\bibfnamefont {G.}~\bibnamefont {Gasbarri}}, \bibinfo
  {author} {\bibfnamefont {A.}~\bibnamefont {Setter}}, \bibinfo {author}
  {\bibfnamefont {C.}~\bibnamefont {Timberlake}}, \ and\ \bibinfo {author}
  {\bibfnamefont {H.}~\bibnamefont {Ulbricht}},\ }\href@noop {} {\bibfield
  {journal} {\bibinfo  {journal} {Physical Review Applied}\ }\textbf {\bibinfo
  {volume} {13}},\ \bibinfo {pages} {064027} (\bibinfo {year}
  {2020})}\BibitemShut {NoStop}%
\bibitem [{\citenamefont {Ahrens}\ \emph {et~al.}(2024)\citenamefont {Ahrens},
  \citenamefont {Ji}, \citenamefont {Budker}, \citenamefont {Timberlake},
  \citenamefont {Ulbricht},\ and\ \citenamefont
  {Vinante}}]{ahrens2024levitated}%
  \BibitemOpen
  \bibfield  {author} {\bibinfo {author} {\bibfnamefont {F.}~\bibnamefont
  {Ahrens}}, \bibinfo {author} {\bibfnamefont {W.}~\bibnamefont {Ji}}, \bibinfo
  {author} {\bibfnamefont {D.}~\bibnamefont {Budker}}, \bibinfo {author}
  {\bibfnamefont {C.}~\bibnamefont {Timberlake}}, \bibinfo {author}
  {\bibfnamefont {H.}~\bibnamefont {Ulbricht}}, \ and\ \bibinfo {author}
  {\bibfnamefont {A.}~\bibnamefont {Vinante}},\ }\href@noop {} {\bibfield
  {journal} {\bibinfo  {journal} {arXiv preprint arXiv:2401.03774}\ } (\bibinfo
  {year} {2024})}\BibitemShut {NoStop}%
\bibitem [{\citenamefont {Simon}\ \emph {et~al.}(2001)\citenamefont {Simon},
  \citenamefont {Heflinger},\ and\ \citenamefont
  {Geim}}]{simon2001diamagnetically}%
  \BibitemOpen
  \bibfield  {author} {\bibinfo {author} {\bibfnamefont {M.}~\bibnamefont
  {Simon}}, \bibinfo {author} {\bibfnamefont {L.}~\bibnamefont {Heflinger}}, \
  and\ \bibinfo {author} {\bibfnamefont {A.}~\bibnamefont {Geim}},\ }\href@noop
  {} {\bibfield  {journal} {\bibinfo  {journal} {American journal of physics}\
  }\textbf {\bibinfo {volume} {69}},\ \bibinfo {pages} {702} (\bibinfo {year}
  {2001})}\BibitemShut {NoStop}%
\bibitem [{\citenamefont {Leng}\ \emph {et~al.}(2024)\citenamefont {Leng},
  \citenamefont {Chen}, \citenamefont {Li}, \citenamefont {Wang}, \citenamefont
  {Wang}, \citenamefont {Wang}, \citenamefont {Xie}, \citenamefont {Duan},
  \citenamefont {Huang},\ and\ \citenamefont {Du}}]{leng2024measurement}%
  \BibitemOpen
  \bibfield  {author} {\bibinfo {author} {\bibfnamefont {Y.}~\bibnamefont
  {Leng}}, \bibinfo {author} {\bibfnamefont {Y.}~\bibnamefont {Chen}}, \bibinfo
  {author} {\bibfnamefont {R.}~\bibnamefont {Li}}, \bibinfo {author}
  {\bibfnamefont {L.}~\bibnamefont {Wang}}, \bibinfo {author} {\bibfnamefont
  {H.}~\bibnamefont {Wang}}, \bibinfo {author} {\bibfnamefont {L.}~\bibnamefont
  {Wang}}, \bibinfo {author} {\bibfnamefont {H.}~\bibnamefont {Xie}}, \bibinfo
  {author} {\bibfnamefont {C.-K.}\ \bibnamefont {Duan}}, \bibinfo {author}
  {\bibfnamefont {P.}~\bibnamefont {Huang}}, \ and\ \bibinfo {author}
  {\bibfnamefont {J.}~\bibnamefont {Du}},\ }\href@noop {} {\bibfield  {journal}
  {\bibinfo  {journal} {Physical Review Letters}\ }\textbf {\bibinfo {volume}
  {132}},\ \bibinfo {pages} {123601} (\bibinfo {year} {2024})}\BibitemShut
  {NoStop}%
\bibitem [{\citenamefont {Perdriat}\ \emph {et~al.}(2023)\citenamefont
  {Perdriat}, \citenamefont {Pellet-Mary}, \citenamefont {Copie},\ and\
  \citenamefont {H{\'e}tet}}]{perdriat2023planar}%
  \BibitemOpen
  \bibfield  {author} {\bibinfo {author} {\bibfnamefont {M.}~\bibnamefont
  {Perdriat}}, \bibinfo {author} {\bibfnamefont {C.}~\bibnamefont
  {Pellet-Mary}}, \bibinfo {author} {\bibfnamefont {T.}~\bibnamefont {Copie}},
  \ and\ \bibinfo {author} {\bibfnamefont {G.}~\bibnamefont {H{\'e}tet}},\
  }\href@noop {} {\bibfield  {journal} {\bibinfo  {journal} {Physical Review
  Research}\ }\textbf {\bibinfo {volume} {5}},\ \bibinfo {pages} {L032045}
  (\bibinfo {year} {2023})}\BibitemShut {NoStop}%
\bibitem [{\citenamefont {Fadeev}\ \emph
  {et~al.}(2021{\natexlab{a}})\citenamefont {Fadeev}, \citenamefont
  {Timberlake}, \citenamefont {Wang}, \citenamefont {Vinante}, \citenamefont
  {Band}, \citenamefont {Budker}, \citenamefont {Sushkov}, \citenamefont
  {Ulbricht},\ and\ \citenamefont {Kimball}}]{fadeev2021ferromagnetic}%
  \BibitemOpen
  \bibfield  {author} {\bibinfo {author} {\bibfnamefont {P.}~\bibnamefont
  {Fadeev}}, \bibinfo {author} {\bibfnamefont {C.}~\bibnamefont {Timberlake}},
  \bibinfo {author} {\bibfnamefont {T.}~\bibnamefont {Wang}}, \bibinfo {author}
  {\bibfnamefont {A.}~\bibnamefont {Vinante}}, \bibinfo {author} {\bibfnamefont
  {Y.~B.}\ \bibnamefont {Band}}, \bibinfo {author} {\bibfnamefont
  {D.}~\bibnamefont {Budker}}, \bibinfo {author} {\bibfnamefont {A.~O.}\
  \bibnamefont {Sushkov}}, \bibinfo {author} {\bibfnamefont {H.}~\bibnamefont
  {Ulbricht}}, \ and\ \bibinfo {author} {\bibfnamefont {D.~F.~J.}\ \bibnamefont
  {Kimball}},\ }\href@noop {} {\bibfield  {journal} {\bibinfo  {journal}
  {Quantum Science and Technology}\ }\textbf {\bibinfo {volume} {6}},\ \bibinfo
  {pages} {024006} (\bibinfo {year} {2021}{\natexlab{a}})}\BibitemShut
  {NoStop}%
\bibitem [{\citenamefont {Fadeev}\ \emph
  {et~al.}(2021{\natexlab{b}})\citenamefont {Fadeev}, \citenamefont {Wang},
  \citenamefont {Band}, \citenamefont {Budker}, \citenamefont {Graham},
  \citenamefont {Sushkov},\ and\ \citenamefont {Kimball}}]{fadeev2021gravity}%
  \BibitemOpen
  \bibfield  {author} {\bibinfo {author} {\bibfnamefont {P.}~\bibnamefont
  {Fadeev}}, \bibinfo {author} {\bibfnamefont {T.}~\bibnamefont {Wang}},
  \bibinfo {author} {\bibfnamefont {Y.}~\bibnamefont {Band}}, \bibinfo {author}
  {\bibfnamefont {D.}~\bibnamefont {Budker}}, \bibinfo {author} {\bibfnamefont
  {P.~W.}\ \bibnamefont {Graham}}, \bibinfo {author} {\bibfnamefont {A.~O.}\
  \bibnamefont {Sushkov}}, \ and\ \bibinfo {author} {\bibfnamefont {D.~F.~J.}\
  \bibnamefont {Kimball}},\ }\href@noop {} {\bibfield  {journal} {\bibinfo
  {journal} {Physical Review D}\ }\textbf {\bibinfo {volume} {103}},\ \bibinfo
  {pages} {044056} (\bibinfo {year} {2021}{\natexlab{b}})}\BibitemShut
  {NoStop}%
\bibitem [{\citenamefont {Ahrens}\ and\ \citenamefont
  {Vinante}(2025)}]{ahrens2025observation}%
  \BibitemOpen
  \bibfield  {author} {\bibinfo {author} {\bibfnamefont {F.}~\bibnamefont
  {Ahrens}}\ and\ \bibinfo {author} {\bibfnamefont {A.}~\bibnamefont
  {Vinante}},\ }\href@noop {} {\bibfield  {journal} {\bibinfo  {journal} {arXiv
  preprint arXiv:2504.13744}\ } (\bibinfo {year} {2025})}\BibitemShut {NoStop}%
\bibitem [{\citenamefont {Domcke}\ \emph {et~al.}(2025)\citenamefont {Domcke},
  \citenamefont {Ellis},\ and\ \citenamefont {Rodd}}]{MagneticWeberBar}%
  \BibitemOpen
  \bibfield  {author} {\bibinfo {author} {\bibfnamefont {V.}~\bibnamefont
  {Domcke}}, \bibinfo {author} {\bibfnamefont {S.~A.~R.}\ \bibnamefont
  {Ellis}}, \ and\ \bibinfo {author} {\bibfnamefont {N.~L.}\ \bibnamefont
  {Rodd}},\ }\href {\doibase 10.1103/966v-r5fm} {\bibfield  {journal} {\bibinfo
   {journal} {Phys. Rev. Lett.}\ }\textbf {\bibinfo {volume} {134}},\ \bibinfo
  {pages} {231401} (\bibinfo {year} {2025})}\BibitemShut {NoStop}%
\bibitem [{\citenamefont {Yadav}\ \emph {et~al.}(2016)\citenamefont {Yadav},
  \citenamefont {Verma},\ and\ \citenamefont {Nagar}}]{yadav2016optimized}%
  \BibitemOpen
  \bibfield  {author} {\bibinfo {author} {\bibfnamefont {S.}~\bibnamefont
  {Yadav}}, \bibinfo {author} {\bibfnamefont {S.}~\bibnamefont {Verma}}, \ and\
  \bibinfo {author} {\bibfnamefont {S.}~\bibnamefont {Nagar}},\ }\href@noop {}
  {\bibfield  {journal} {\bibinfo  {journal} {Ifac-PapersOnLine}\ }\textbf
  {\bibinfo {volume} {49}},\ \bibinfo {pages} {778} (\bibinfo {year}
  {2016})}\BibitemShut {NoStop}%
\bibitem [{\citenamefont {El~Hajjaji}\ and\ \citenamefont
  {Ouladsine}(2002)}]{el2002modeling}%
  \BibitemOpen
  \bibfield  {author} {\bibinfo {author} {\bibfnamefont {A.}~\bibnamefont
  {El~Hajjaji}}\ and\ \bibinfo {author} {\bibfnamefont {M.}~\bibnamefont
  {Ouladsine}},\ }\href@noop {} {\bibfield  {journal} {\bibinfo  {journal}
  {IEEE Transactions on industrial Electronics}\ }\textbf {\bibinfo {volume}
  {48}},\ \bibinfo {pages} {831} (\bibinfo {year} {2002})}\BibitemShut
  {NoStop}%
\bibitem [{\citenamefont {Saitoh}\ \emph {et~al.}(2020)\citenamefont {Saitoh},
  \citenamefont {Stoneking},\ and\ \citenamefont
  {Pedersen}}]{saitoh2020levitated}%
  \BibitemOpen
  \bibfield  {author} {\bibinfo {author} {\bibfnamefont {H.}~\bibnamefont
  {Saitoh}}, \bibinfo {author} {\bibfnamefont {M.}~\bibnamefont {Stoneking}}, \
  and\ \bibinfo {author} {\bibfnamefont {T.~S.}\ \bibnamefont {Pedersen}},\
  }\href@noop {} {\bibfield  {journal} {\bibinfo  {journal} {Review of
  Scientific Instruments}\ }\textbf {\bibinfo {volume} {91}} (\bibinfo {year}
  {2020})}\BibitemShut {NoStop}%
\bibitem [{\citenamefont {Morikawa}\ \emph {et~al.}(2004)\citenamefont
  {Morikawa}, \citenamefont {Ohkuni}, \citenamefont {Hori}, \citenamefont
  {Yamakoshi}, \citenamefont {Goto}, \citenamefont {Ogawa}, \citenamefont
  {Yanagi},\ and\ \citenamefont {Mito}}]{morikawa2004plasma}%
  \BibitemOpen
  \bibfield  {author} {\bibinfo {author} {\bibfnamefont {J.}~\bibnamefont
  {Morikawa}}, \bibinfo {author} {\bibfnamefont {K.}~\bibnamefont {Ohkuni}},
  \bibinfo {author} {\bibfnamefont {D.}~\bibnamefont {Hori}}, \bibinfo {author}
  {\bibfnamefont {S.}~\bibnamefont {Yamakoshi}}, \bibinfo {author}
  {\bibfnamefont {T.}~\bibnamefont {Goto}}, \bibinfo {author} {\bibfnamefont
  {Y.}~\bibnamefont {Ogawa}}, \bibinfo {author} {\bibfnamefont
  {N.}~\bibnamefont {Yanagi}}, \ and\ \bibinfo {author} {\bibfnamefont
  {T.}~\bibnamefont {Mito}},\ }\href@noop {} {\bibfield  {journal} {\bibinfo
  {journal} {Teion Kogaku}\ }\textbf {\bibinfo {volume} {39}} (\bibinfo {year}
  {2004})}\BibitemShut {NoStop}%
\bibitem [{\citenamefont {Lotz}(2022)}]{Lotz2022}%
  \BibitemOpen
  \bibfield  {author} {\bibinfo {author} {\bibfnamefont {C.}~\bibnamefont
  {Lotz}},\ }\href@noop {} {\bibfield  {journal} {\bibinfo  {journal} {Doctoral
  dissertation}\ } (\bibinfo {year} {2022})}\BibitemShut {NoStop}%
\bibitem [{SM()}]{SM}%
  \BibitemOpen
  \href@noop {} {\enquote {\bibinfo {title} {Supplementary materials},}\
  }\bibinfo {note} {Materials and methods, supplementary text, figures, and
  tables are available as supplementary materials.}\BibitemShut {Stop}%
\bibitem [{\citenamefont {Liang}\ \emph {et~al.}(2018)\citenamefont {Liang},
  \citenamefont {De-Sheng}, \citenamefont {Wei-Biao}, \citenamefont {Tang},
  \citenamefont {Qiu-Zhi}, \citenamefont {Bin}, \citenamefont {Lin},
  \citenamefont {Wei}, \citenamefont {Zuo-Ren},\ and\ \citenamefont
  {Jian-Bo}}]{2018In}%
  \BibitemOpen
  \bibfield  {author} {\bibinfo {author} {\bibfnamefont {L.}~\bibnamefont
  {Liang}}, \bibinfo {author} {\bibfnamefont {L.}~\bibnamefont {De-Sheng}},
  \bibinfo {author} {\bibfnamefont {C.}~\bibnamefont {Wei-Biao}}, \bibinfo
  {author} {\bibfnamefont {L.}~\bibnamefont {Tang}}, \bibinfo {author}
  {\bibfnamefont {Q.}~\bibnamefont {Qiu-Zhi}}, \bibinfo {author} {\bibfnamefont
  {W.}~\bibnamefont {Bin}}, \bibinfo {author} {\bibfnamefont {L.}~\bibnamefont
  {Lin}}, \bibinfo {author} {\bibfnamefont {R.}~\bibnamefont {Wei}}, \bibinfo
  {author} {\bibfnamefont {D.}~\bibnamefont {Zuo-Ren}}, \ and\ \bibinfo
  {author} {\bibfnamefont {Z.}~\bibnamefont {Jian-Bo}},\ }\href@noop {}
  {\bibfield  {journal} {\bibinfo  {journal} {Nature Communications}\ }\textbf
  {\bibinfo {volume} {9}},\ \bibinfo {pages} {2760} (\bibinfo {year}
  {2018})}\BibitemShut {NoStop}%
\bibitem [{\citenamefont {Cacciapuoti}\ \emph {et~al.}(2024)\citenamefont
  {Cacciapuoti}, \citenamefont {Busso}, \citenamefont {Jansen}, \citenamefont
  {Pataraia}, \citenamefont {Peignier}, \citenamefont {Weinberg}, \citenamefont
  {Crescence}, \citenamefont {Helm}, \citenamefont {Kehrer}, \citenamefont
  {Koller} \emph {et~al.}}]{cacciapuoti2024atomic}%
  \BibitemOpen
  \bibfield  {author} {\bibinfo {author} {\bibfnamefont {L.}~\bibnamefont
  {Cacciapuoti}}, \bibinfo {author} {\bibfnamefont {A.}~\bibnamefont {Busso}},
  \bibinfo {author} {\bibfnamefont {R.}~\bibnamefont {Jansen}}, \bibinfo
  {author} {\bibfnamefont {S.}~\bibnamefont {Pataraia}}, \bibinfo {author}
  {\bibfnamefont {T.}~\bibnamefont {Peignier}}, \bibinfo {author}
  {\bibfnamefont {S.}~\bibnamefont {Weinberg}}, \bibinfo {author}
  {\bibfnamefont {P.}~\bibnamefont {Crescence}}, \bibinfo {author}
  {\bibfnamefont {A.}~\bibnamefont {Helm}}, \bibinfo {author} {\bibfnamefont
  {J.}~\bibnamefont {Kehrer}}, \bibinfo {author} {\bibfnamefont
  {S.}~\bibnamefont {Koller}},  \emph {et~al.},\ }in\ \href@noop {} {\emph
  {\bibinfo {booktitle} {Journal of Physics: Conference Series}}},\ Vol.\
  \bibinfo {volume} {2889}\ (\bibinfo {organization} {IOP Publishing},\
  \bibinfo {year} {2024})\ p.\ \bibinfo {pages} {012005}\BibitemShut {NoStop}%
\bibitem [{\citenamefont {Aveline}\ \emph {et~al.}(2020)\citenamefont
  {Aveline}, \citenamefont {Williams}, \citenamefont {Elliott}, \citenamefont
  {Dutenhoffer}, \citenamefont {Kellogg}, \citenamefont {Kohel}, \citenamefont
  {Lay}, \citenamefont {Oudrhiri}, \citenamefont {Shotwell}, \citenamefont {Yu}
  \emph {et~al.}}]{aveline2020observation}%
  \BibitemOpen
  \bibfield  {author} {\bibinfo {author} {\bibfnamefont {D.~C.}\ \bibnamefont
  {Aveline}}, \bibinfo {author} {\bibfnamefont {J.~R.}\ \bibnamefont
  {Williams}}, \bibinfo {author} {\bibfnamefont {E.~R.}\ \bibnamefont
  {Elliott}}, \bibinfo {author} {\bibfnamefont {C.}~\bibnamefont
  {Dutenhoffer}}, \bibinfo {author} {\bibfnamefont {J.~R.}\ \bibnamefont
  {Kellogg}}, \bibinfo {author} {\bibfnamefont {J.~M.}\ \bibnamefont {Kohel}},
  \bibinfo {author} {\bibfnamefont {N.~E.}\ \bibnamefont {Lay}}, \bibinfo
  {author} {\bibfnamefont {K.}~\bibnamefont {Oudrhiri}}, \bibinfo {author}
  {\bibfnamefont {R.~F.}\ \bibnamefont {Shotwell}}, \bibinfo {author}
  {\bibfnamefont {N.}~\bibnamefont {Yu}},  \emph {et~al.},\ }\href@noop {}
  {\bibfield  {journal} {\bibinfo  {journal} {Nature}\ }\textbf {\bibinfo
  {volume} {582}},\ \bibinfo {pages} {193} (\bibinfo {year}
  {2020})}\BibitemShut {NoStop}%
\bibitem [{\citenamefont {Elliott}\ \emph {et~al.}(2018)\citenamefont
  {Elliott}, \citenamefont {Krutzik}, \citenamefont {Williams}, \citenamefont
  {Thompson},\ and\ \citenamefont {Aveline}}]{elliott2018nasa}%
  \BibitemOpen
  \bibfield  {author} {\bibinfo {author} {\bibfnamefont {E.~R.}\ \bibnamefont
  {Elliott}}, \bibinfo {author} {\bibfnamefont {M.~C.}\ \bibnamefont
  {Krutzik}}, \bibinfo {author} {\bibfnamefont {J.~R.}\ \bibnamefont
  {Williams}}, \bibinfo {author} {\bibfnamefont {R.~J.}\ \bibnamefont
  {Thompson}}, \ and\ \bibinfo {author} {\bibfnamefont {D.~C.}\ \bibnamefont
  {Aveline}},\ }\href@noop {} {\bibfield  {journal} {\bibinfo  {journal} {npj
  Microgravity}\ }\textbf {\bibinfo {volume} {4}},\ \bibinfo {pages} {16}
  (\bibinfo {year} {2018})}\BibitemShut {NoStop}%
\bibitem [{\citenamefont {Li}\ \emph {et~al.}(2025)\citenamefont {Li},
  \citenamefont {Chen}, \citenamefont {Zhang}, \citenamefont {Wang},
  \citenamefont {Zhou}, \citenamefont {He}, \citenamefont {Fang}, \citenamefont
  {Zhou}, \citenamefont {He}, \citenamefont {Jiang} \emph
  {et~al.}}]{li2025realization}%
  \BibitemOpen
  \bibfield  {author} {\bibinfo {author} {\bibfnamefont {J.}~\bibnamefont
  {Li}}, \bibinfo {author} {\bibfnamefont {X.}~\bibnamefont {Chen}}, \bibinfo
  {author} {\bibfnamefont {D.}~\bibnamefont {Zhang}}, \bibinfo {author}
  {\bibfnamefont {W.}~\bibnamefont {Wang}}, \bibinfo {author} {\bibfnamefont
  {Y.}~\bibnamefont {Zhou}}, \bibinfo {author} {\bibfnamefont {M.}~\bibnamefont
  {He}}, \bibinfo {author} {\bibfnamefont {J.}~\bibnamefont {Fang}}, \bibinfo
  {author} {\bibfnamefont {L.}~\bibnamefont {Zhou}}, \bibinfo {author}
  {\bibfnamefont {C.}~\bibnamefont {He}}, \bibinfo {author} {\bibfnamefont
  {J.}~\bibnamefont {Jiang}},  \emph {et~al.},\ }\href@noop {} {\bibfield
  {journal} {\bibinfo  {journal} {National Science Review}\ }\textbf {\bibinfo
  {volume} {12}},\ \bibinfo {pages} {nwaf012} (\bibinfo {year}
  {2025})}\BibitemShut {NoStop}%
\bibitem [{\citenamefont {He}\ \emph {et~al.}(2023)\citenamefont {He},
  \citenamefont {Chen}, \citenamefont {Fang}, \citenamefont {Chen},
  \citenamefont {Sun}, \citenamefont {Wang}, \citenamefont {Zhong},
  \citenamefont {Zhou}, \citenamefont {He}, \citenamefont {Li} \emph
  {et~al.}}]{he2023space}%
  \BibitemOpen
  \bibfield  {author} {\bibinfo {author} {\bibfnamefont {M.}~\bibnamefont
  {He}}, \bibinfo {author} {\bibfnamefont {X.}~\bibnamefont {Chen}}, \bibinfo
  {author} {\bibfnamefont {J.}~\bibnamefont {Fang}}, \bibinfo {author}
  {\bibfnamefont {Q.}~\bibnamefont {Chen}}, \bibinfo {author} {\bibfnamefont
  {H.}~\bibnamefont {Sun}}, \bibinfo {author} {\bibfnamefont {Y.}~\bibnamefont
  {Wang}}, \bibinfo {author} {\bibfnamefont {J.}~\bibnamefont {Zhong}},
  \bibinfo {author} {\bibfnamefont {L.}~\bibnamefont {Zhou}}, \bibinfo {author}
  {\bibfnamefont {C.}~\bibnamefont {He}}, \bibinfo {author} {\bibfnamefont
  {J.}~\bibnamefont {Li}},  \emph {et~al.},\ }\href@noop {} {\bibfield
  {journal} {\bibinfo  {journal} {npj Microgravity}\ }\textbf {\bibinfo
  {volume} {9}},\ \bibinfo {pages} {58} (\bibinfo {year} {2023})}\BibitemShut
  {NoStop}%
\end{thebibliography}

%

\clearpage

\end{document}